\documentclass[sigconf]{acmart}

\usepackage{multicol,graphicx,mathrsfs,amsmath,pifont,amscd,latexsym,color,fancyhdr,url,longtable, pifont, subfigure,epstopdf,setspace,multirow,colortbl,tabularx, booktabs, bbm,listings, balance,color,indentfirst,algorithmic,algorithm,enumitem}

\begin{document}

\copyrightyear{2018} \acmYear{2018} \setcopyright{acmcopyright} \acmConference[WSDM 2018]{WSDM 2018: The Eleventh ACM International Conference on Web Search and Data Mining }{February 5--9, 2018}{Marina Del Rey, CA, USA} \acmBooktitle{WSDM 2018: The Eleventh ACM International Conference on Web Search and Data Mining , February 5--9, 2018, Marina Del Rey, CA, USA} \acmPrice{15.00} \acmDOI{10.1145/3159652.3159690} \acmISBN{978-1-4503-5581-0/18/02}
% \fancyhead{} 

% Title portion
\title[A Deep Learning Framework for News-oriented Stock Trend Prediction]{Listening to Chaotic Whispers: A Deep Learning Framework for News-oriented Stock Trend Prediction}

\author{Ziniu Hu}
\authornote{This work was done when the author was an intern at Microsoft Research Asia.}
\affiliation{%
  \department{Key Lab of High-Confidence Software Technology, MoE}
  \institution{Peking University, China}
}
\email{bull@pku.edu.cn}

\author{Weiqing Liu}
\affiliation{%
  \institution{Microsoft Research}
  \city{Beijing}
  \country{China}}
\email{Weiqing.Liu@microsoft.com}

\author{Jiang Bian}
\affiliation{%
  \institution{Microsoft Research}
  \city{Beijing}
  \country{China}}
\email{Jiang.Bian@microsoft.com}

\author{Xuanzhe Liu}
\affiliation{%
  \department{Key Lab of High-Confidence Software Technology, MoE}
  \institution{Peking University, China}}
\email{liuxuanzhe@pku.edu.cn}

\author{Tie-Yan Liu}
\affiliation{%
  \institution{Microsoft Research}
  \city{Beijing}
  \country{China}}
\email{Tie-Yan.Liu@microsoft.com}

\begin{abstract}
Stock trend prediction plays a critical role in seeking maximized profit from the stock investment. However, precise trend prediction is very difficult since the highly volatile and non-stationary nature of the stock market. Exploding information on the Internet together with the advancing development of natural language processing and text mining techniques have enabled investors to unveil market trends and volatility from online content. Unfortunately, the quality, trustworthiness, and comprehensiveness of online content related to stock market vary drastically, and a large portion consists of the low-quality news, comments, or even rumors. To address this challenge, we imitate the learning process of human beings facing such chaotic online news, driven by three principles: {\em sequential content dependency}, {\em diverse influence}, and {\em effective and efficient learning}. In this paper, to capture the first two principles, we designed a Hybrid Attention Networks (HAN) to predict the stock trend based on the sequence of recent related news. Moreover, we apply the self-paced learning mechanism to imitate the third principle. Extensive experiments on real-world stock market data demonstrate the effectiveness of our framework. A further simulation illustrates that a straightforward trading strategy based on our proposed framework can significantly increase the annualized return. 
\end{abstract}

%\begin{CCSXML}
%<ccs2012>
%<concept>
%<concept_id>10010147.10010178.10010179.10003352</concept_id>
%<concept_desc>Computing methodologies~Information extraction</concept_desc>
%<concept_significance>300</concept_significance>
%</concept>
%</ccs2012>
%\end{CCSXML}

%\ccsdesc[300]{Computing methodologies~Information extraction}


% % We no longer use \terms command
% \terms{Design, Algorithms, Performance}

\keywords{stock trend prediction; deep learning; text mining}

\maketitle
%\footnotetext[1]{This work was done when the author was an intern at Microsoft Research Asia.}

\section{Introduction}\label{sec:introduction}

In order for seeking maximized profit, stock investors continuously attempt to predict the future trends of market \cite{Nuij2014An}, which, however, is quite challenging due to highly volatile and non-stationary nature of the market \cite{adam2016stock}. Traditional efforts on predicting the stock trend have been carried out based on information from various fields. One of most basic ways relies on technical analysis upon recent prices and volumes on the market. Such methods yield the very limitations on unveiling the rules that govern the drastic dynamics of the market. Meanwhile, another basic method focuses on analyzing financial statements of each company, which is though incapable of catching the impact of recent trends. 

With the rapid growth of the Internet, content from online media has indeed become a gold mine for investors to understand market trends and volatility. Even more, the advancing development of Natural Language Processing techniques has inspired increasing efforts on stock trend prediction by automatically analyzing stock-related articles. For instance, Tetlock et al. \cite{Tetlock2007Giving} extracts and quantifies the optimism and pessimism of Wall Street Journal reports and observes that trading volume tends to increase after pessimism reports and high pessimism scored reports tend to be followed by a downtrend and a reversion of market prices.

Not surprisingly, the effectiveness of such textual analyses depends on the quality of target articles. For instance, comparing with reading a comprehensive report of a company from Wall Street, analyzing a simple declarative news about this company is less likely to produce an accurate prediction. Unfortunately, the quality, trustworthiness, and comprehensiveness of online content related to stock market vary drastically, and a large portion of the online content consists of the low-quality news, comments, or even rumors. 

To address this challenge, we imitate the learning process of human beings facing such chaotic online news, which can be summarized into three principles:
\begin{itemize}[wide, leftmargin = 0.5pc]
\item \textbf{Sequential Context Dependency}: Even if a single piece of news is very likely to be of low-quality or not informative enough, human can comprehensively consider a sequence of related recent news as a unified context of the stock and consequently make a more reliable prediction on the subsequent stock trend. More importantly, within the unified sequential context, humans can pay different attention to various parts according to their respective importance and influence.

\item \textbf{Diverse Influence}: While the influence of different online news can be diverse greatly, human beings can discriminate them based on the intrinsic content. For instance, some breaking news (e.g. air crash, military conflict) will profoundly affect the trend of related stocks, whilst some useless comments or vague rumors may cause little disturb on the stock trends. In real-world investment, people tend to consciously and comprehensively consider the estimated impact of each news at the time of predicting the subsequent stock trend.

\item \textbf{Effective and Efficient Learning}: News cannot always provide obviously informative indications on stock trend, especially when the content of news contains vague information on stock trend or even there is very limited number of news about certain stocks in a period. In order for both effective and efficient learning, human beings tend to first gain knowledge by focusing on informative occasions, and then turn to disturbing evidence to obtain uneasy experience.
\end{itemize}

To capture the first two principles of human learning process, we design a Hybrid Attention Networks (HAN) to predict the stock trend based on the sequence of recent related news. First, to imitate the human cognition on sequential context with diverse attentions, we construct attention-based recurrent neural networks (RNN) at the higher level. In particular, the RNN structure enables the processing of recent related news for a stock in a unified sequence, and the attention mechanism is capable of identifying more influential time periods of the sequence. Second, to further model diverse influence of news, we propose news-level attention-based neural networks at the lower level, which aims at recognizing more important news from others at the same time point. 

To imitate the effective and efficient learning of human, we employ the self-paced learning (SPL) \cite{Kumar2011Self} mechanism. Since the news-based stock trend prediction is more challenging in some situations, SPL enables us to automatically skip those training samples from some challenging periods in the early stage of model training, and progressively increase the complexity of training samples. Self-paced learning mechanism can automatically choose the suitable training samples for different training stage, which enhanced the final performance of our framework.

To validate the effectiveness of our approach, in this paper, we performed extensive experiments on real-world data. Comparing with traditional approaches, the experiment results show that our framework can significantly improve the performance of stock trend prediction. Furthermore, we simulated the stock investment using a simple trading strategy based on our framework, and the results illustrate that a straightforward trading strategy can achieve much better annualized return than the baseline methods.

To sum up, the contributions of our work include: 
\begin{itemize}
\item A summarization of principles for imitating the learning process of human beings, particularly for stock trend prediction from the chaotic online news.
\item A Hybrid Attention Networks with self-paced learning for stock trend prediction, driven by principles of human learning process. 
\item Experimental studies on real-world data with simulated investment performance based on the real stock market. 
\end{itemize}

The rest of the paper is organized as follows. We introduce related work in Section 2. We present empirical analysis to reveal principles for designing news-oriented stock prediction framework in Section 3, based on which we propose a new deep learning framework with details in Section 4. Experimental setup and results are demonstrated in Section 5. We conclude the paper and point out future directions in Section 6.

\section{Related Work}\label{sec:related}
Stock trend prediction has attracted many research efforts due to its decisive role in stock investment. In general, traditional approaches can be categorized into two primary approaches: technical and fundamental analysis, according to the various types of information they mainly relied on. 

Technical analysis deals with the time-series historic market-data, such as trading price and volume, and make predictions based on that. The main goal of this type of approach is to discover the trading patterns that can be leveraged for future prediction. One of the most widely used model in this direction is the Autoregressive (AR) model for linear and stationary time-series~\cite{Li2016AR}. However, the non-linear and non-stationary nature of stock prices limits the applicality of AR models. Hence, previous studies attempted to applied non-linear learning methods~\cite{Nayak2015SVMKNN} to catch the complex patterns underlying the market trend. With the developement of deep learning, more research efforts have been paid on exploiting deep neurual networks for financial prediction~\cite{Kim2012simultaneous,Laboissiere2015maximum,Gocken2016IMA,Adebiyi2014Comparison,Patel2015predicting,Ticknor2013bayesian}. To further model the long-term dependency in time series, recurrent neurual networks (RNN), especially Long Short-Term Memory (LSTM) network, have also been employed in financial predition~\cite{Rather2015recurrent,Akita2016deep,Gao2016stock}. In most recent time, Zhang~\textit{et al.}~\cite{Zhang2017MultiFreq} proposed a new RNN, called State Frequency Memory (SFM), to discover multi-frequency trading patterns for stock price prediction.

One major limitation of technical analysis is that it is incapable of unveiling the rules that govern the dynamics of the market beyond price data. Fundamental approaches, on the contrary, seek information from outside market-historic-data, such as geopolitics, financial environment and business principles. Information explosion on the Internet has promoted online content, especially news, as one of the most important sources for fundamental analysis.
There have been many attempts to mine news data for better predicting market trends. Nassirtoussi \textit{et al.}~\cite{Nassirtoussi2015Text} proposed a multi-layer dimension reduction algorithm with semantics and sentiment to predict intraday directional-movements of a currency-pair in the foreign exchange market. Ding \textit{et al.}~\cite{Ding2015Deep} proposed a deep learning method for event-driven stock market prediction. They further augmented their approach~\cite{ding2016knowledge} by incorporating an outside knowledge graph into the learning process for event embeddings. Wang \textit{et al.}~\cite{Wang2014A} performed a text regression task to predict the volatility of stock prices. Xie \textit{et al.}~\cite{Xie2013Semantic} introduced a novel tree representation, and use it to train predictive models with tree kernels using support vector machines. Hagenau \textit{et al.}~\cite{Hagenau2013Automated} extract a large scale of expressive features to represent the unstructured text data and employs a robust feature selection to enhance the stock prediction.

Another major aspects for market news mining is to analyze sentiments from public news and social media, and then use it to predict market trends. Li \textit{et al.}~\cite{Li2014News} implements a generic stock price prediction framework using sentiment analysis. Zhou\textit{et al.} \cite{zhou2016can} studies particularly the Chinese stock market. They conduct a thorough study over 10 million stock-relevant tweets from Weibo, and find five attributes that stock market in China can be competently predicted by various online emotions. Nguyen \textit{et al.}\cite{si2013exploiting} explicitly consider the topics relating to the target stocks, and  extracting topics and related sentiments from social media to make prediction.

While there have been many efforts in exploiting news for stock prediction, few of them paid enough attention to the quality, trustworthiness, and comprehensiveness of news, which highly affects the effectiveness of textual analysis. In this paper, we address the challenge of chaotic news by imitating the learning process of human beings, inspired by which we propose a Hybrid Attention Networks (HAN) to predict the stock trend based on the sequence of recent related news and employ self-paced learning for effective and efficient learning.

\section{Empirical Analysis}\label{sec:analysis}

\begin{figure*}[t]
\centering
\includegraphics[width=0.9\textwidth]	
{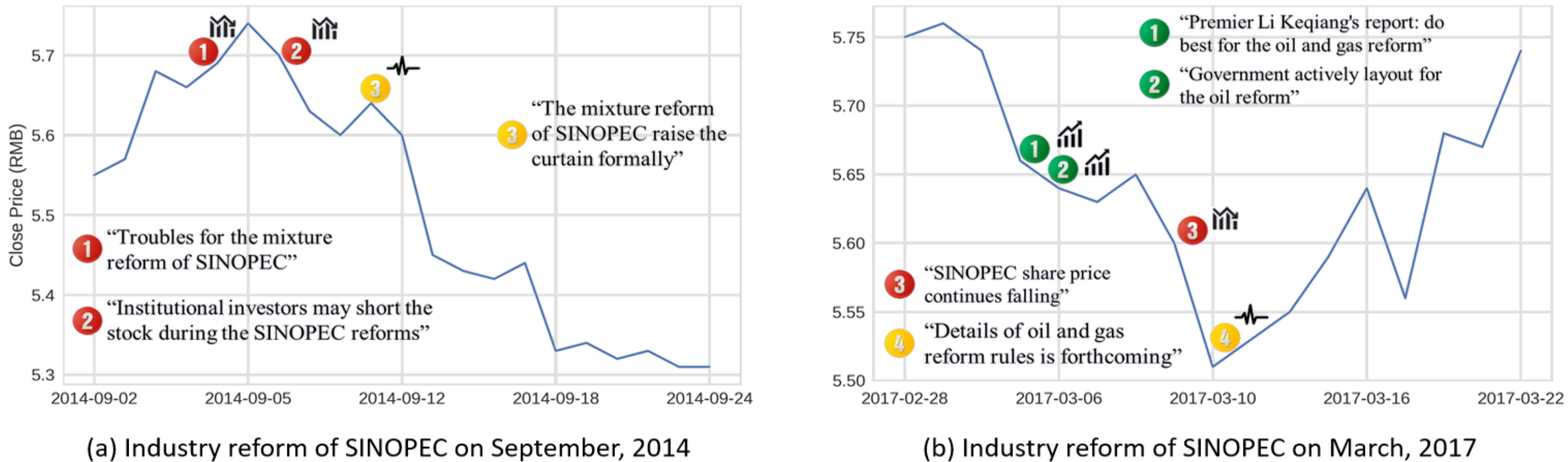}
\caption{An example of two news sequences about the petrol industry reform, jointly shown with the stock price of SINOPEC that is one of the biggest petrol companies in China. In the figure, red, green and yellow circles represent negative, positive and neutral news respectively.}
\label{fig:exp} 
\end{figure*}

\begin{figure}[ht]
\centering
\includegraphics[width=0.48\textwidth]  
  {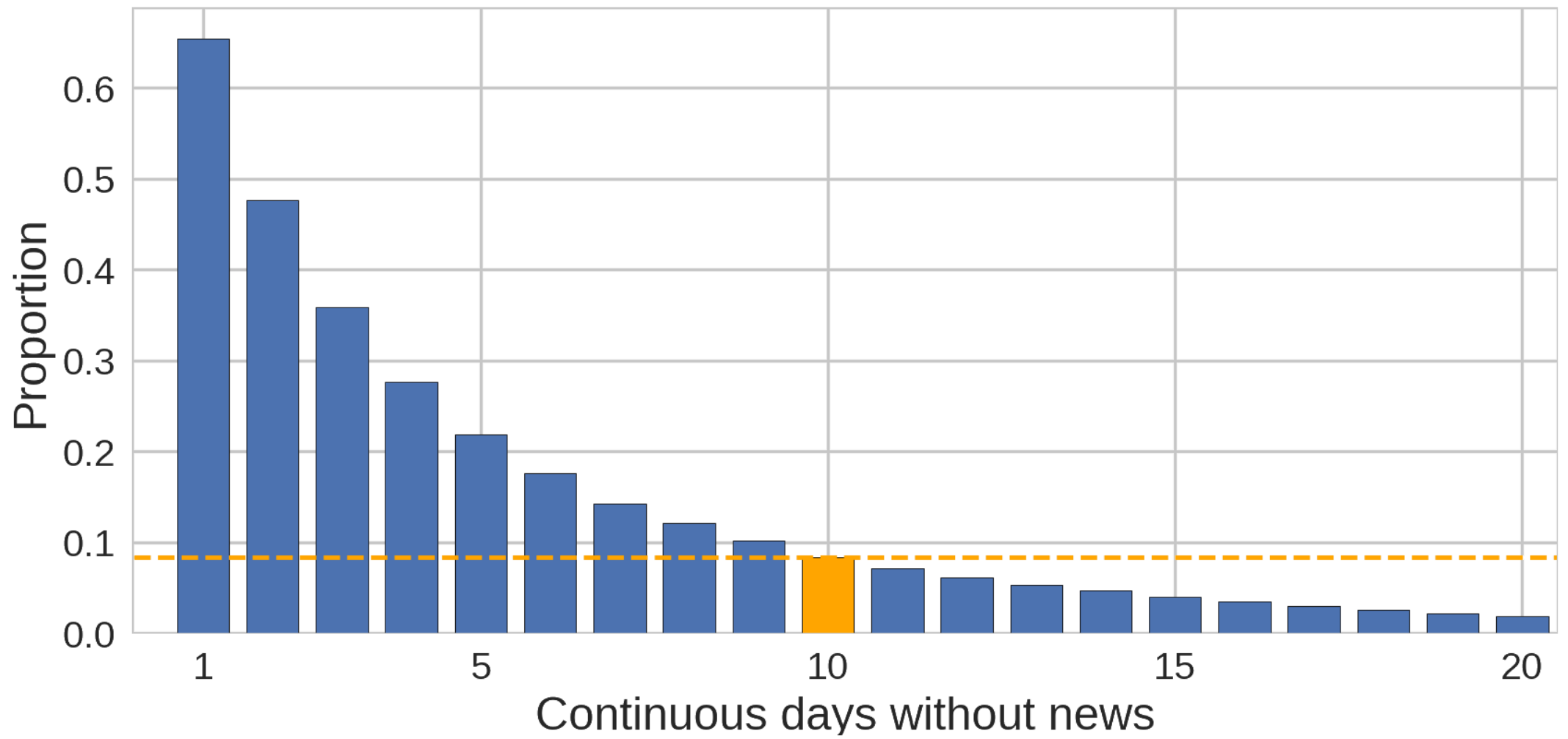}
\caption{Proportion of the consecutive days without news.}
\label{fig:cdf}
\end{figure}

In this section, through empirical analysis, we will reveal three principles of the human learning process with respect to stock trend prediction via chaotic news. These principles can consequently provide essential guidelines in designing our learning framework.

\subsection{Sequential Context Dependency}

Due to the diverse quality of online financial news, human investors usually prefer not to rely on a single piece of news to make prediction due to its limited or even vague information. Instead, by broadly analyzing a sequence of news and combining them into a unified context, each news can provide complementary information and thus a more reliable assessment of stock trend can be made. 

For example, Figure~\ref{fig:exp} illustrates two news sequences referring to petrol industry reforms happening in September 2014 and March 2017, respectively. The figure also displays the share price of SINOPEC, which is one of the biggest petrol companies in China.

From the figure, we can see two yellow circles with the \textit{fluctuation} sign that represent two news declaring the initiation of two reforms, respectively. By merely referring to the two news, we can hardly tell the future trend of SINOPEC, since they revealed quite limited details about the reforms. However, the difference between these two reforms could be inferred by their previous news sequences. In particular, previous news with the \textit{down} sign on September 2014 indicated that the reform might cause negative effects; while on March 2017, the news with the \textit{rise} sign demonstrates quite positive signals. 

In reality, human investors can naturally synthesize these analytical reports before the reforms actually began, in order to better assess the influence of these reforms on relevant stocks. Therefore, to imitate such analysis process as a human, an ideal framework should integrate and interpret each news in a sequential temporal context, rather than analyze them separately.

\subsection{Diverse Influence}
 
Significant news has more intensive and durable influence on the market than those trivial ones. For example, the third news with the \textit{down} sign in Figure~\ref{fig:exp}(b) summarizes that the share price of SINOPEC has been continuously going down, which should give a negative signal to the stock trend. However, compared with the positive news reporting the reform on petrol industry in the same period, this negative one yields much less importance. As it turns out later, after the SINOPEC's price drops down for only one day, it starts to rise up and keeps the uptrend for quite a long time, proving that the influence of the negative news is indeed weaker than the positive one.

Based on the diversity of news influence, we can conclude that when analyzing the news, an ideal framework should have the ability to distinguish the news with more intensive and durable influence, and pay more attention to them.

\subsection{Effective and Efficient Learning}~\label{sec:eelrn}

%Since the stock trend is determined by various factors, including investors' sentiment, event triggering, price momentum, etc. The news source can only contain a partial information of the market. Therefore, sometimes the news are predictive to the stock trend while on other occasions the news are less informative, especially those times when the news amount is too scarce to make credible prediction. 

News cannot always provide an informative indication of the stock trend, especially when there exists only an insufficient number of news about specific stocks in a period.
%It is challenging to predict the trend of a stock which has no news for a period of time.
According to the online news we collected, Figure~\ref{fig:cdf} shows the occurrence rates of the situation that there is no news about a particular stock ``Jiai Technology'' for consecutive $l$ days or more.
Within those $8.4\%$ of time periods when there is no news reported for more than $10$ days, it is quite tough to make any news-oriented prediction.
Additionally, other situations, such as the aggregation of vague news, also introduce difficulties to the prediction.
%Figure~\ref{fig:cdf} shows that for any given length $l$ of days, the number of temporal sequence of a length $l$ without news divided by the total number of sequence with the length of $l$.
%From the figure, we can see that there are about $23.1\%$ sequences of the length as $5$ without news, and about $9.2\%$ sequences of the length as $20$ without news. 
%This result shows that in our dataset, there exist a lot of sequences which doesn't contain any or only contains scarce news, by which the prediction can be extremely difficult. 

%In order for both effective and efficient learning, human beings tend to first gain knowledge by focusing on informative occasions, and then turn to disturbing evidences to obtain uneasy experience.
Such diverse difficulty in news-oriented stock trend prediction task has motivated human investors to find a more effective and efficient learning process. In reality, human investors tend to first gain an overall knowledge by focusing on common occasions, and then turn to exceptional cases. 

Inspired by this, an ideal learning framework for stock prediction should follow a similar process, which in particular conducts learning on more informative news at the earlier stage, and further optimized to tackle harder samples.

% A straightforward approach is to filter out the training data with scarce news input with heuristic rules. Recently, a novel learning theory called Self-Paced learning. This approach is inspired by the underlying cognitive processes of humans and animals, which generally start with learning easier aspects of a task, and then gradually take more complex examples into consideration. In self-paced learning, the model learner must determine what is easy versus complex estimated by error metric, and assign sample importance by this metric. To alleviate the side effect of scarce news data and train an adaptive model, we adopt the self-paced learning technique to train our model.

%\section{Problem Statement}\label{sec:problem}
%\input{section/problem}

\section{Deep Learning Framework for News-Oriented Trend  Prediction}\label{sec:approach}
% In this section, we present our noise-adaptive framework, which incorporating an attention network to rank each news's reliability and then integrate these news to make prediction. The attention network and the prediction network can be trained end-to-end. Also, we incorporate a curriculum mechanism which feeds the model with the "easier" data first.

In this section, we first formalize the problem of the stock trend prediction. Then, we present our framework based on the three design principles discussed in Empirical Analysis (Section~\ref{sec:analysis}). We first propose a Hybrid Attention Networks (HAN), which consists of two attention layers on news level and temporal level, respectively. Next, we incorporate a self-paced learning mechanism that enables the model to adjust the learning sequence in order to achieve better performance.

\subsection{Problem Statement}

We regard the problem of stock trend prediction as a classification problem. For a given date $t$ and a given stock $s$, 
%we can calculate the rise percent of the stock price compared to the next date $t+1$ by:
we can calculate its rise percent by:
\begin{displaymath}
Rise\_Percent(t) = \frac{Open\_Price(t+1) - Open\_Price(t)} {Open\_Price(t)}
\end{displaymath}

Similar to many previous studies, we can divide rise percent into three classes: \textit{DOWN}, \textit{UP}, and \textit{PRESERVE}, representing the significant dropping, rising, and steady stock trend on the next date, respectively.  

The stock trend prediction task can be formulated as follows: given the length of a time sequence $N$, the stock $s$ and date $t$, the goal is to use the news corpus sequence from time $t-N$ to $t-1$, denoted as $[C_{t-N}, C_{t-N+1}, ..., C_{t-1}]$, to predict the class of $Rise\_Percent(t)$, i.e. \textit{DOWN}, \textit{UP}, or \textit{PRESERVE}. Note that each news corpus $C_i$ contains a set of news with the size of $L$, $C_i = [n_{i1}, n_{i2}, ..., n_{iL}]$, denoting $L$ related news on date $i$.
%The task is to design a model to classify the $Rise\_Label(t)$, indicating the market trend of the next time stamp $t$. 
% Intuitively, the influence of news on stock trend is similar among different stocks. Therefore, we simply build one single classifier to accomplish this task.

\subsection{Hybrid Attention Networks}

Based on the \textbf{Sequential Context Dependency} principle, our framework should interpret and analyze news in the sequential temporal context and pay more attention to critical time periods. In addition, based on the \textbf{Diverse Influence} principle, our framework should distinguish more significant news from others. To capture these two principles, we design a hybrid attention network (HAN), which incorporates attention mechanisms at both the news level and the temporal level. 

\begin{figure}[t]
\centering   
\includegraphics [width=0.5\textwidth]{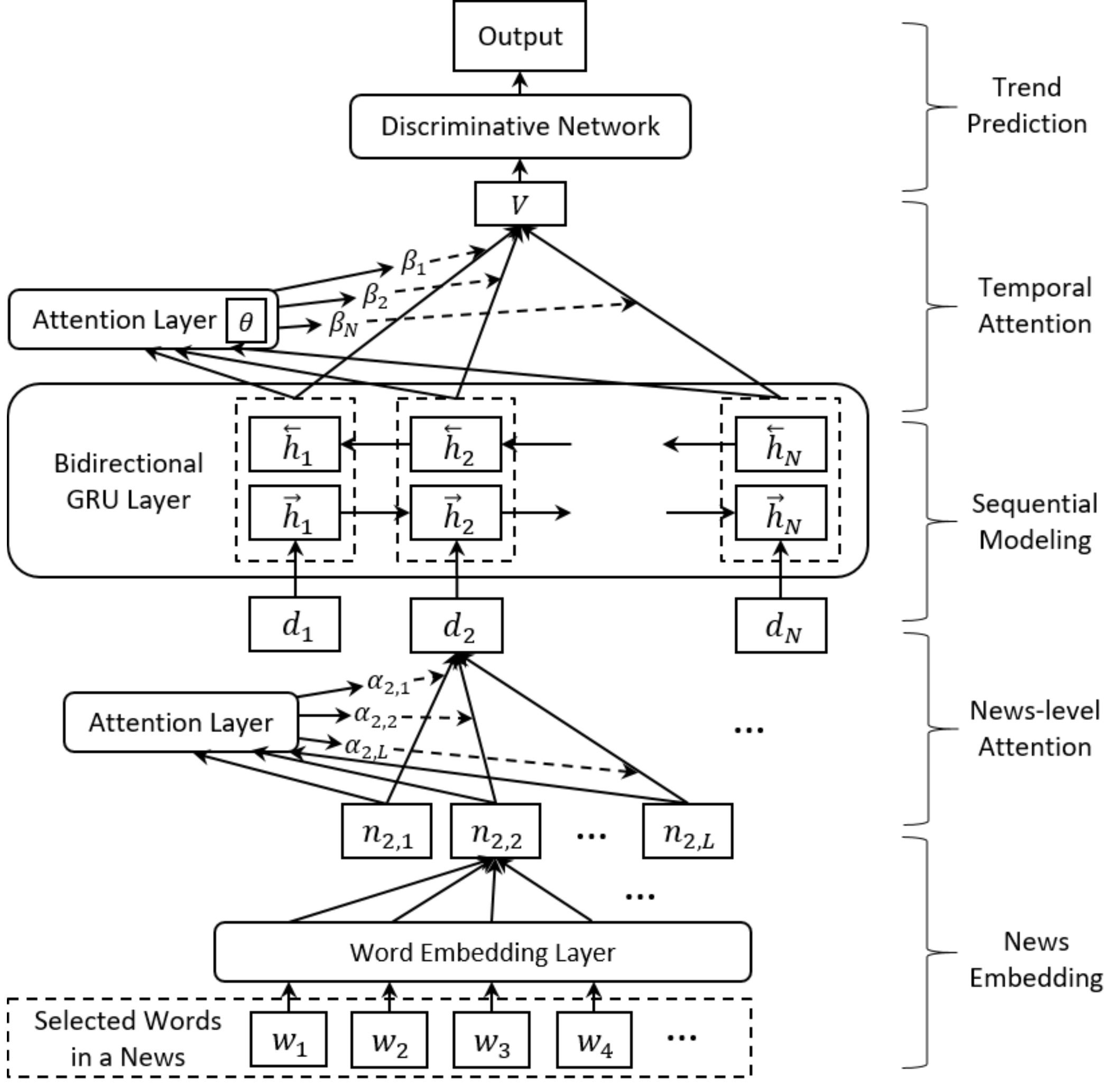} 
\caption{The overall framework of the Hybrid Attention Networks (HAN).} 
 \label{fig:arch} 
\end{figure}

We summarize the overall framework in Figure~\ref{fig:arch}. Given the input of a news corpus sequence, a news embedding layer encodes each news into a news vector $n_{ti}$. Next, a news-level attention layer assigns an attention value to each news vector in a date, and calculate the weighted mean of these news vectors as a corpus vector for this date. Afterwards, these corpus vectors are encoded by a bi-directional Gated Recurrent Units (GRU). Then, another temporal attention layer assigns an attention value to each date, and calculate the weighted mean of these encoded corpus vectors to represent the overall sequential context information. Finally, the classification is made by a discriminative network. The details of the architecture are elaborated below.

% Inspired by the work of \cite{Yang2017Hierarchical}, we use an Attention-based recurrent neural network organized in a hierarchical structure to model the financial news sequence

% Based on our first assumption, that the news should be analyzed and interpreted in a sequential context, we adopt a recurrent neural network to model the daily news sequence. In addition, based on the second assumption, not all documents are equally significant for stock trend prediction, and the detection of noisy documents should involve both its own content and the context documents in the same time period. Therefore, we build two hierarchical-attention layers to distinguish news with different importance. On the lower layer, every news in a date are gathered as input, and 

\noindent\textbf{News Embedding:}
For each $i^{th}$ news in news corpus $C_t$ of date $t$, we use a word embedding layer to calculate the embedded vector for each word and then average all the words' vectors to construct a news vector $n_{ti}$. To reduce the complexity of the framework, we pre-train an unsupervised Word2Vec as the word embedding layer rather than tuning its parameters in the learning process.

\noindent\textbf{News-level Attention:}
Since not all news contributes equally to predicting the stock trend, we introduce an attention mechanism to aggregate the news weighted by an assigned attention value, in order to reward the news offering critical information. Specifically,
\begin{equation*}
\begin{aligned}
& u_{ti} = sigmoid(W_n n_{ti} + b_n)\\
& \alpha_{ti} = \frac{exp(u_{ti})}{\sum_j exp(u_{tj})}\\
& d_t = \sum_i \alpha_{ti} n_{ti}
\end{aligned}
\end{equation*}

We first estimate attention values by feeding the news vector $n_{ti}$ through a one-layer network to get the news-level attention value $u_{ti}$, and then calculate a normalized attention weight $\alpha_{ti}$ through a softmax function. Finally, we calculate the overall corpus vector $d_t$ as a weighted sum of each news vector respectively, and use this vector to represent all news information for date $t$. Thus, we get a temporal sequence of corpus vector $D = [d_i], i \in [1,N]$. Obviously, the attention layer can be trained end-to-end and thus gradually learn to assign more attention to the reliable and informative news based on its content.

\noindent\textbf{Sequential Modeling:}
To encode the temporal sequence of corpus vectors, we adopt Gated Recurrent Units (GRU). %, as the document encoder. 
GRU is a variant of recurrent neural networks that uses a gating mechanism to check the state of sequences without separate memory cells. At date $t$, the GRU computes the news state $h_t$ by linearly interpolating the previous state $h_{t-1}$ and the current updated state $\tilde{h_t}$, as:

$$h_t = (1-z_t) * h_{t-1} + z_t * \tilde{h_t}$$

The current updated state $\tilde{h_t}$ is computed by non-linearly combining the corpus vector input for this time-stamp and the previous state, as:

$$\tilde{h_t} = tanh(W_h d_t + r_t * (U_h h_{t-1}) + b_h)$$

\noindent where $r_t$ denotes the reset gate, controlling how much past state should be used for updating the new state, and $z_t$ is the update gate, deciding how much past information should be kept and how much new information should be added. These two gates are calculated by:
\begin{equation*}
\begin{aligned}
r_t = \sigma(W_r d_t + U_r h_{t-1} + b_r)\\
z_t = \sigma (W_z d_t + U_z h_{t-1} + b_z)\\
\end{aligned}
\end{equation*}

Therefore, we can get the latent vector for each date $t$ through GRU. In order to capture the information from the past and future of a news as its context, we concatenate the latent vectors from both directions to construct a bi-directional encoded vector $h_i$ as:
\begin{equation*}
\begin{aligned}
& \overrightarrow{h_i} = \overrightarrow{GRU}(d_i), i \in [1,L]\\
& \overleftarrow{h_i} = \overleftarrow{GRU}(d_i), i \in [L,1]\\
& h_i = [\overrightarrow{h_i} , \overleftarrow{h_i}]\\
\end{aligned}
\end{equation*}

The resulted $h_i$ incorporates the information of both its surrounding context and itself. In this way, we encode the temporal sequence of corpus vectors.

\noindent\textbf{Temporal Attention:}
% Since not all documents contribute equally to predict the stock trend. In order to reward those documents that offer clues to correctly classify the market state, we introduce attention mechanism to aggregate the documents weighted by an assigned attention value. Specifically,
Since the news published at different dates contribute to the stock trend unequally, we adopt the temporal-level attention mechanism, which incorporates both the inherent temporal pattern and the news content, to distinguish the temporal difference as:
\begin{equation*}
\begin{aligned}
& o_i = sigmoid(W_h h_i + b_h)\\
& \beta_i = \frac{exp(\theta_i o_i)}{\sum_j exp(\theta_j o_j)}\\
& V = \sum_i \beta_i h_i
\end{aligned}
\end{equation*}

\noindent where $\theta_i$ is the parameter for each date in the softmax layer, indicating in general which date is more significant, and $o$ is the latent representations of encoded corpus vectors. By combining them through a softmax layer, we can get an attention vector $\beta$ to distinguish the temporal difference. Then we use $\beta$ to calculate the weighted sum $V$, so that it can incorporate the sequential news context information with temporal attention, and will be used for classification.

\noindent\textbf{Trend Prediction:}
The final discriminative network is a standard Multi-layer Perceptron (MLP), which takes $V$ as input and produces the three-class classification of the future stock trend.

\subsection{Self-paced Learning Mechanism}

As discussed in Section~\ref{sec:eelrn}, there exist some natural challenges of news-oriented stock trend prediction, such as the scarceness of news and the aggregation of vague news, which cause severe learning difficulties. To conduct an effective and efficient learning, the designed model should skip those challenging training samples at the early training stages, and progressively incorporate them into the model training.

%Due to the non-convex nature of the document classification problem and the complexity of a RNN model, it takes a particularly long time for our proposed model to converge, and the result is not promising. 
%it is challenging to train an optimal model.
%Moreover, some situations introduced more difficulties in the model training, such as the scarceness of news discussed in Section~\ref{sec:analysis}.3.

%Here, we propose to make the training process more effective by using the concept of 
Curriculum Learning~\cite{Bengio2009Curriculum} is a learning mode that can imitate such a learning process.
%. The key idea of Curriculum Leaning is to 
%can guide the model training by carefully choosing samples that are simple for model to learn, and progressively increase the complexity of training samples. 
%The curriculum determines a static sequence of training samples and is often derived by predetermined heuristics. 
%Although heuristic knowledge often proves to be useful, 
However, the sequence of training samples in curriculum learning is fixed by predetermined heuristics (curriculum), which 
%the fixed curriculum 
cannot be adjusted to the feedback from the dynamic learned models. To alleviate this issue, Kuma \textit{et al.}~\cite{Kumar2011Self} designed %a new paradigm called 
Self-Paced Learning (SPL) to embed curriculum design into the learning objective, so it can jointly optimize the curriculum and the learned model simultaneously.

%There exist many sequences without any news, which are definitely difficult samples to learn. This is an example of difficult samples, while there also may exist some other cases. 
%Since we cannot enumerate all the heuristic rules for the stock trend problem, we propose to employ self-paced learning that lets the model automatically learn the learning curriculum itself.

%To leverage the side effect of training difficulty, we use the SPL into our framework so that the it 
Therefore, we take advantage of SPL in our framework to
learn the news influence in an organized manner. Formally, given a training set $D = {(x_i, y_i)}_{i=1}^n$, where $x_i \in R^m$ denotes all the news inputs for the $i^{th}$ observed sample, and $y_i$ represents the corresponding stock trend label. Let $L(y_i, HAN(x_i,w))$ denotes the loss function between label $y_i$ and the output of the whole model $HAN(x_i,w)$, %where $HAN$ denotes the attention-based recurrent network discussed before, 
and $w$ represents the model parameter to be learned. We assign each learning sample an importance weight $v_i$. The goal of SPL is to jointly learn the model parameter $w$ and the latent weight $v = [v1,...,v_n]$ by:
\begin{equation}\label{eq:learn} 
\min_{w,v \in [0,1]^n} E(w,v,\lambda)=\sum_{i=1}^n v_i L(y_i, HAN(x_i,w)) + f(v;\lambda) 
\end{equation}
where $f$ denotes a self-paced regularizer which controls the learning scheme for penalizing the latent weight variables, and $\lambda$ is a hyper-parameter that controls the pace at which the model learns new samples. In this paper, we choose the linear regularizer proposed in ~\cite{Jiang2015Self}, which linearly discriminates samples with respect to their loss, as:
\begin{equation}\label{eq:reg} 
f(v;\lambda) = \frac{1}{2} \lambda \sum_{i=1}^n (v_i^2 - 2v_i)
\end{equation}

Similar to~\cite{Jiang2015Self}~\cite{Kumar2011Self}, we adopt Alternative Convex Search (ACS)\cite{gorski2007biconvex} to solve Equation~\eqref{eq:learn}. 
%Equation\ref{eq:learn}. 
ACS divides the variables into two disjoint blocks. In each iteration, a block of variables is optimized while fixing the other block. With the fixed $w$, the unconstrained close-formed solution for the linear regularizer \eqref{eq:reg} can be calculated by:
%regularizer\ref{eq:reg}.
\begin{equation*}
v_i^*= arg min_v E(w^*,v,\lambda) =
\left\{
\begin{aligned}
-\frac{1}{\lambda} l_i +1 & & l_i < \lambda \\
0 & & 						  l_i \geq \lambda \\ 
\end{aligned} 
\right.
\label{eq:opt} 
\end{equation*}
where $v_i^*$ denotes the $i^{th}$ element in the iterated optimal solution, and $l_i$ denotes the loss for each element $L(y_i, HAN(x_i,w))$. In this way, the latent weight for samples that are different to what model has already learned will receive a linear penalty.

\begin{algorithm}[htb] 
\caption{ Self-paced Learning.} 
\label{alg:Framwork} 
\begin{algorithmic}[1] 
\REQUIRE  
Input dataset, $D$; 
linear regularizer, $f$;
self-paced step size, $\mu$; 
\ENSURE  
Model Parameter, $w$; 
\STATE  Initialize $v^*$ evenly; 
\WHILE {not converged}
	\STATE  Update $w^* = arg min_w  E(w,v^*,\lambda)$;
    \STATE  Update $v^* = arg min_v  E(w^*,v,\lambda)$;
    \IF {$\lambda$ is small}
    	\STATE  increase $\lambda$ by the stepsize $\mu$;
	\ENDIF   
\ENDWHILE 
\RETURN $w^*$; 
\end{algorithmic} 
\end{algorithm}

Based on the optimization equation, the overall learning process can be abstracted as Algorithm \ref{alg:Framwork}. It alternates between two optimization steps described in ACS until the model finally converges: Step 3 learns the optimal model parameters with the fixed and most recent $v^*$ by standard back-propagation mechanism; Step 4 learns the optimal weight variables with the fixed $w^*$ by the linear regulizer. Noted that a small initial $\lambda$ let only samples with rather small loss affect the learning. As the training progresses and $\lambda$ increases in Step 6, samples with larger loss will be gradually incorporated, realizing the effective and efficient learning. 

% If we keep increasing $\lambda$, the model will ultimately use the whole dataset for training, including those noisy data, which will obviously diminish the model performance. To this end, we adopt a early stopping mechanism by stopping the increase of $\lambda$ after a certain number of iterations, in order to filter out those noisy data. The exact stopping iteration threshold is a hyper-parameter of our framework and can be tuned on a small validation set.

\section{Evaluation}\label{sec:evaluation}
\begin{figure}[t]
\centering   
\includegraphics [width=0.40\textwidth]{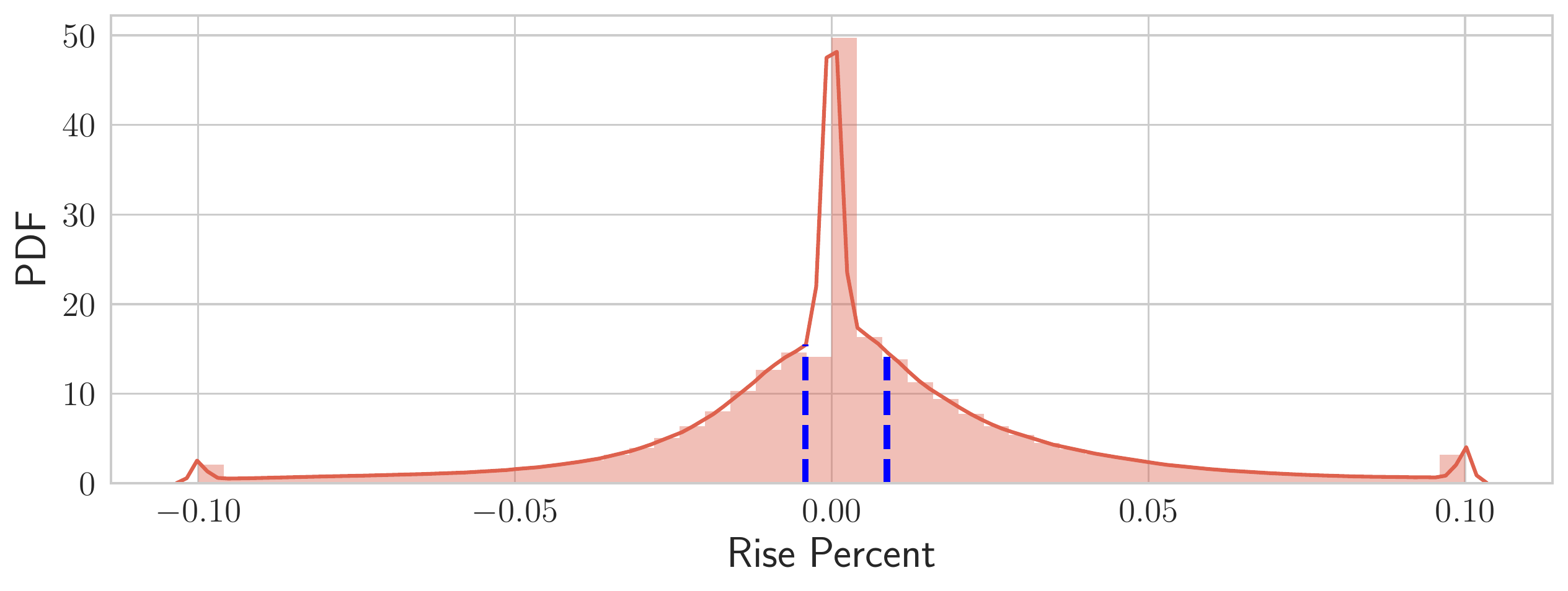} 
\caption{Distribution of the rise percent among the whole dataset.} 
 \label{fig:dist} 
\end{figure}

\begin{figure*}[t]
\centering   
\includegraphics [width=1\textwidth]{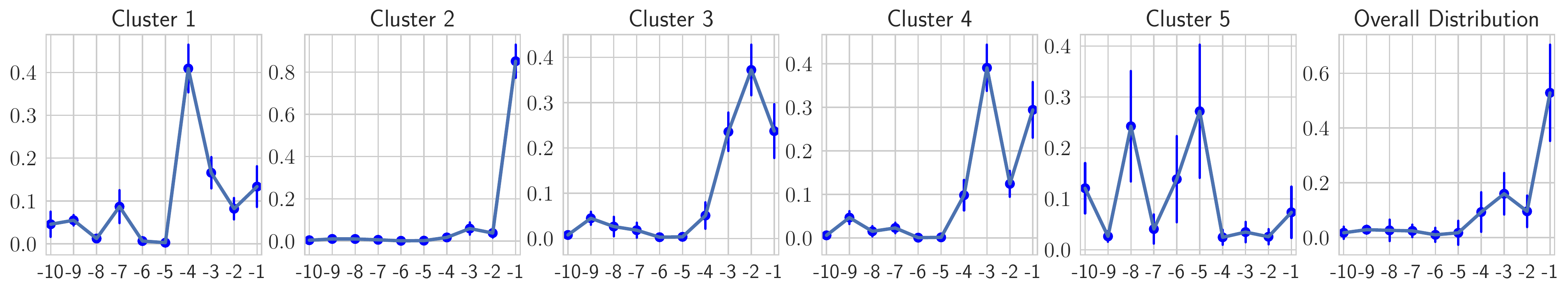} 
\caption{The five attention weight clusters for temporal attention in different context, along with an overall distribution of all the attention weights.} 
 \label{fig:cluster} 
\end{figure*}
In this section, we first present our experimental setup. Then, we conduct comprehensive experiments to evaluate the performance of our proposed deep learning framework, followed by a systematic trading simulation to examine the effectiveness of our framework on the real-world market.

\begin{figure}[!ht]
\centering   
\includegraphics [width=0.45\textwidth]{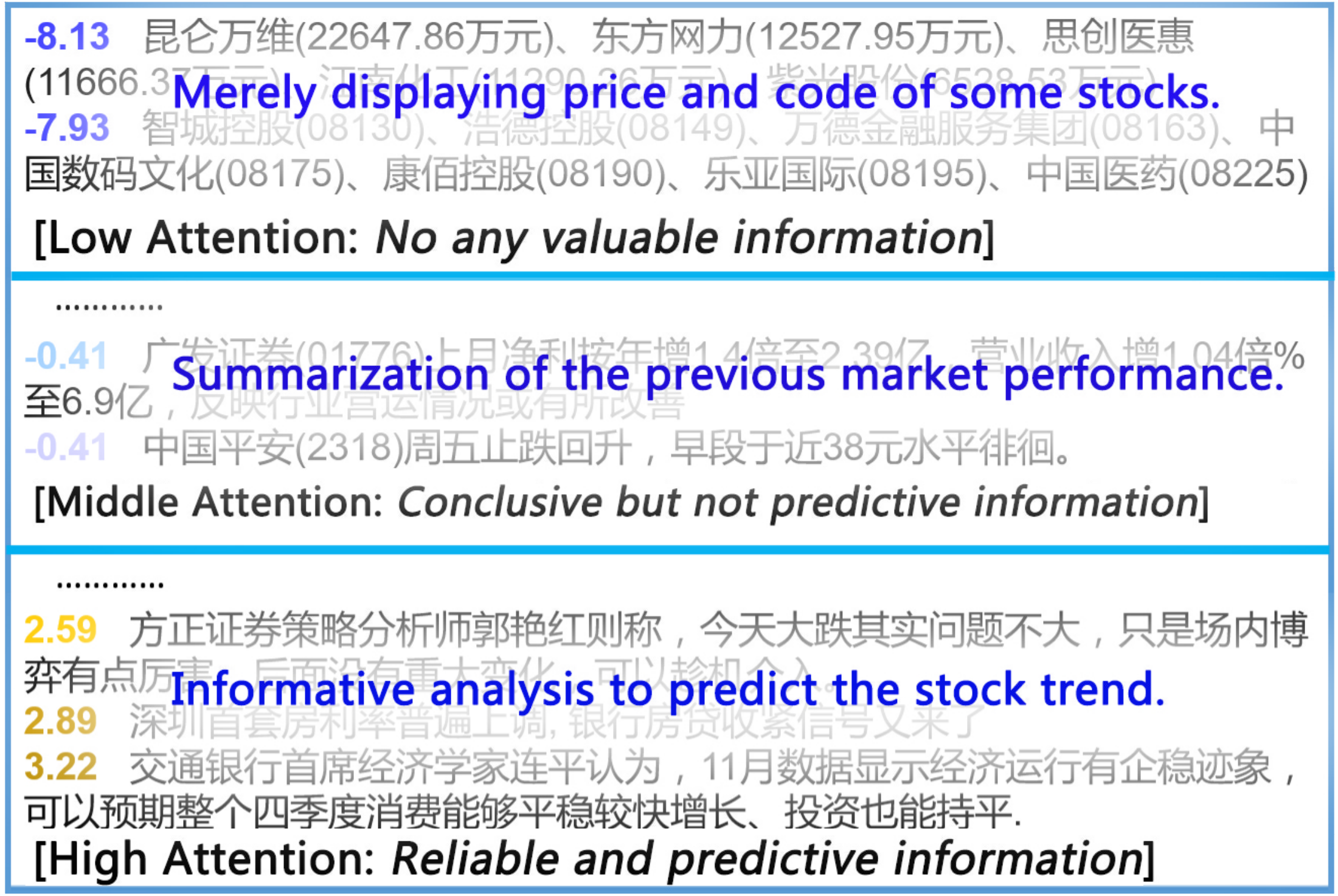} 
\caption{Three classes of demonstrative news in different levels of attention values.} 
 \label{fig:example} 
\end{figure}

\begin{figure}[t]
\centering   
\includegraphics [width=0.40\textwidth]{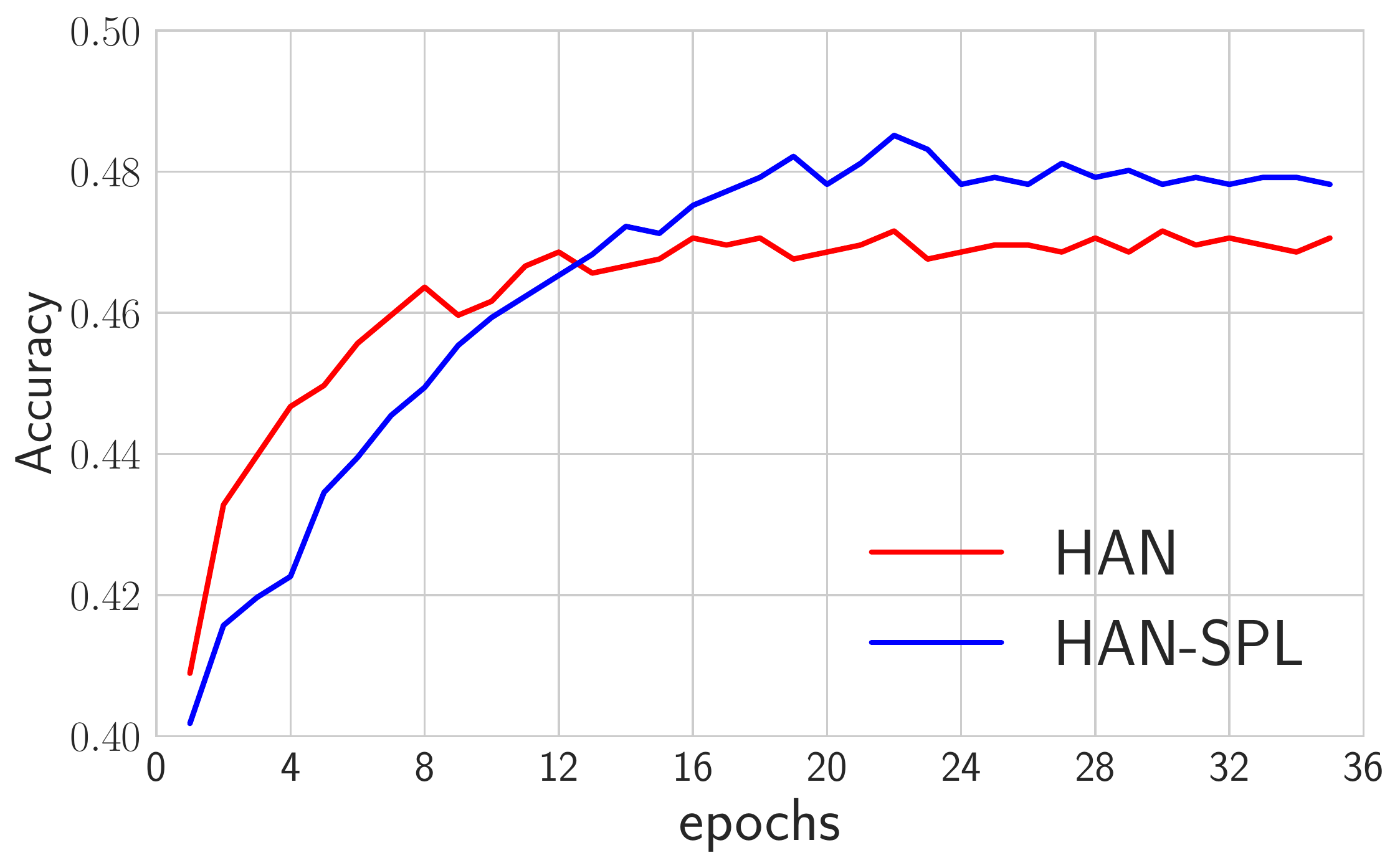} 
\caption{The learning curves of the model using self-paced learning and that without using SPL.} 
 \label{fig:curve} 
\end{figure}

\subsection{Experimental Setup}

\subsubsection{Data Collection}

We obtained the Chinese stock data, including time series in terms of price and trading volume from 2014 to 2017 in daily frequency. There are totally 2527 stocks, covering the vast majority of Chinese stocks.

In the meantime, we collected $1,271,442$ economic news between 2014 and 2017\footnote{The news data are collected from two famous economic websites in China: ~\url{http://www.eastmoney.com/} and ~\url{http://finance.sina.com.cn/}.}. For each news, we extracted the publication time-stamp, title, and its content. 

Then, we correlate each of the collected news to a specific stock if the news mentioned the name of the stock in the title or content. We then filter out the news without any correlation to stocks. After such a process, the total news amount is $425,250$. For each stock, we then aggregate all the news in a certain date to construct the daily news corpus.
%For each stock $s$, we only collect the news sentences that contains the stock name and regarded them as the news sentences related to $s$. 
%Such simple mapping method may undoubtedly skip a lot of informative news which does not contain any stock names but are highly related to them, but it can filter irrelevant news to avoid noise, which is significant for our later analysis. We can use some finer-grained preprocessing methods in the future work.
%

%denoted as $C_t= [n_{t1}, n_{t2}, ..., n_{tL}]$, where %$C_t$ denotes the news corpus at time-stamp $t$, and 
%$n_{ti}$ denotes the $i^{th}$ news in this corpus $C_t$.
\subsubsection{Learning Settings}

To specify the label of the classification problem, we set up two particular thresholds to bin the rise percents, i.e., \textit{DOWN} ($Rise\_Percent(t) < -0.41\%$), \textit{UP} ($Rise\_Percent(t) > 0.87\%$), and \textit{PRESERVE} ($-0.41\% \leq Rise\_Percent(t) \leq 0.87\%$). We define the thresholds so that the three categories are approximately even, as is shown in Figure~\ref{fig:dist}. 

For the following experiments, we split the dataset into a training set ($66.7\%$) from September 2014 to May 2016, and a test set ($33.3\%$) from May 2016 to March 2017. Then we further randomly sample a validation set from the training set with $10\%$ size of it, in order to optimize the hyper-parameters and choose the best epoch.

In all the following experiments, we tokenize each news and remove the stop words and the words appearing less than 5 times to build the vocabulary. We then obtain the word-level embedding by training an un-supervised CBOW Word2Vec\cite{mikolov2013efficient} model with the dimension of 500 through the whole dataset. All the models use the length of a time sequence $N$ as 10.

\begin{figure*}
\begin{minipage}[t]{0.5\linewidth}
\centering   
\includegraphics [width=0.9\textwidth]{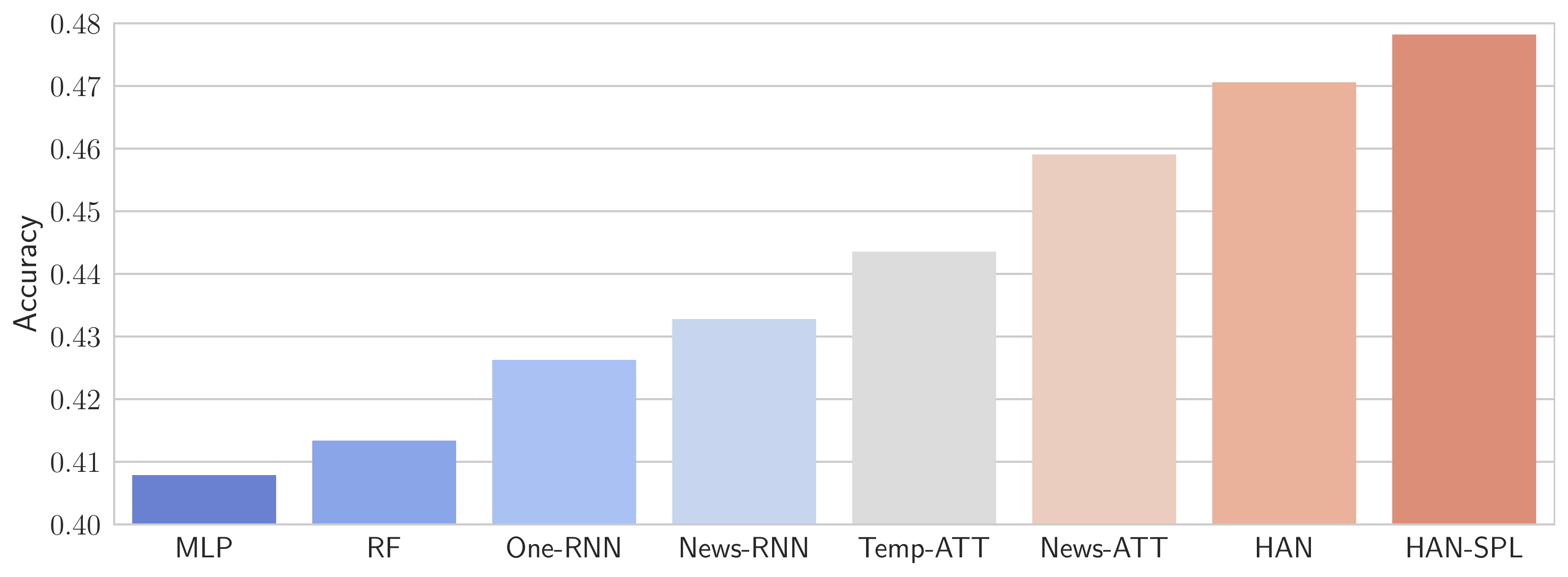} 
\caption{The accuracy result for different methods.} 
 \label{fig:result}
\end{minipage}%
\begin{minipage}[t]{0.5\linewidth}
\centering   
\includegraphics [width=0.9\textwidth]{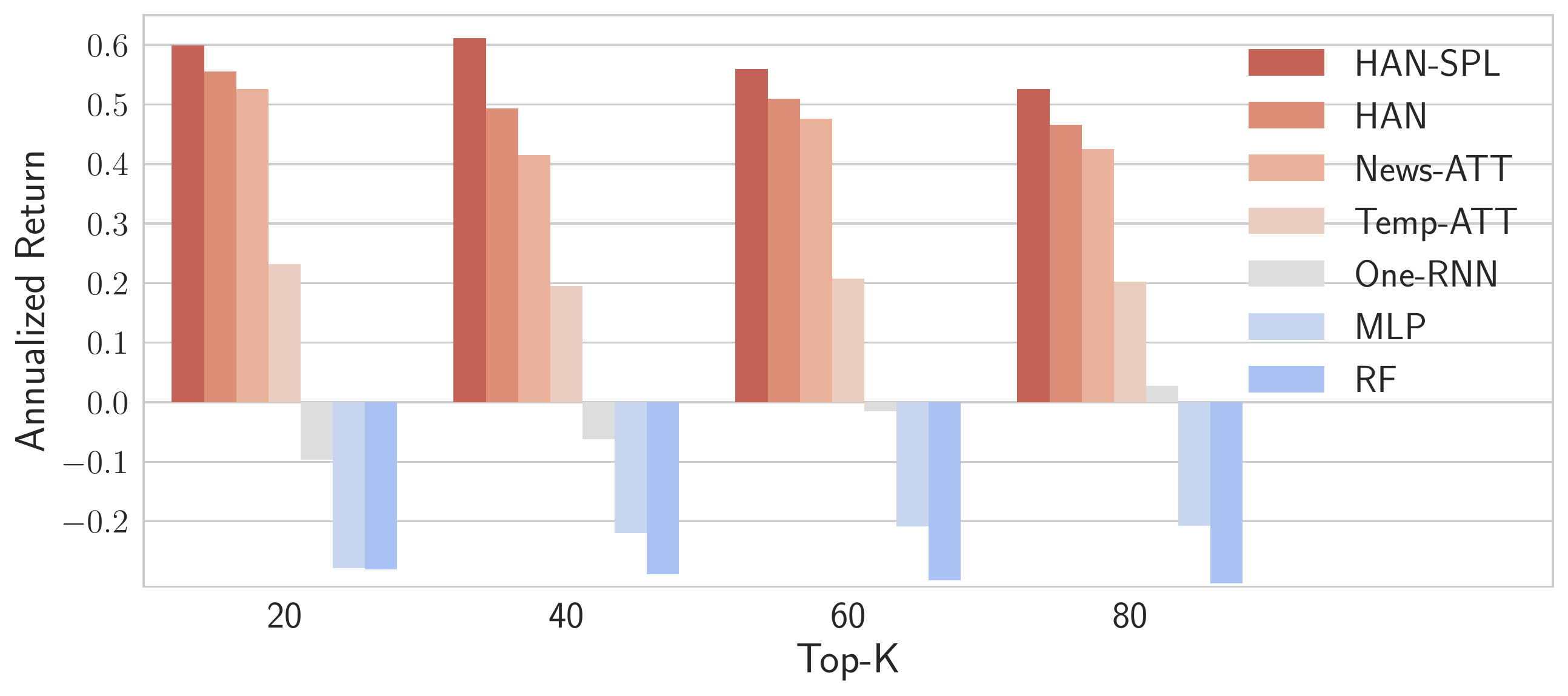} 
\caption{The annualized return for different methods.} 
 \label{fig:annu} 
\end{minipage}
\end{figure*}
\subsubsection{Compared Methods}

To evaluate the effectiveness of our proposed deep learning framework, we compare our result with the performance of the following methods:

\noindent\textbf{Random Forest}: we use the Random Forest (RF) classifier, with the number of trees in the forest as $200$. we define a corpus vector for one date by averaging all news vectors of a certain date, and concatenate the corpus vector in 10 days to construct the input.

\noindent\textbf{Multi-layer Perceptron}: we use the multilayer perceptron (MLP) classifier, with three layers of sizes as 256, 128, and 64, respectively. The input of MLP is the same as that in RF.

\noindent\textbf{News-RNN}: We use a bi-directional GRU-based RNN, with the dimensionality of RNN output layer as 64. The RNN model takes corpus vectors organized in a temporal sequence as the input.

\noindent\textbf{One-RNN}: To evaluate the effectiveness of bi-directional setting, we use a standard one-direction GRU-based RNN as a comparison with News-RNN.

\noindent\textbf{Temporal-Attention-RNN}: To evaluate the temporal attention layer, we use Temporal-Attention-RNN (Temp-ATT) by adding a single temporal-level attention layer on top of the News-RNN.

\noindent\textbf{News-Attention-RNN}: To evaluate the news attention layer, we use News-Attention-RNN (News-ATT) by adding a single news-level attention layer to process corpus vector before the News-RNN.
%The input format of this baseline is the same as our proposed HAN, which has already discussed above.

\noindent\textbf{HAN}: Our proposed hybrid attention networks with the normal learning process.

\noindent\textbf{HAN-SPL}: Our proposed hybrid attention networks with self-paced learning process.

%Also, we compare the two classification results of our proposed HAN with and without using self-paced learning mechanism, in order to evaluate its effectiveness.

%by further dividing the training set into $90\%$ for model fitting and $10\%$ for parameter validation. 

\subsection{Effects of Two Attention Mechanisms}

\subsubsection{Demonstration on News Attention}

Since the news attention values are constructed by a single perception with the embedded news vector as input, we can calculate the attention values for every piece of news on the testing dataset. To validate whether our framework can select significant news and filter out those less informative ones, Figure~\ref{fig:example} shows some detailed results of the news attention. In this figure, the number in front of the news is the attention value, and it shows demonstrative news within three classes that have the lowest, middle, and highest attention values, respectively. From the figure, we can find that the sentences with the lowest attention value merely display the price and the code of the corresponding stock, which obviously have no implication to the future trend; the news with the middle attention value records the previous market performance of the corresponding stock, which provides some information of the stock;
%, but less predictive than the news relating the events or analysis. 
the news with the highest value describes some prediction from official analyst or some significant events, which explicitly contain the predictive information to the market trend. These cases clearly illustrate that our news attention mechanism can indeed distinguish important news from uninformative ones, explaining why it can boost the performance.

\subsubsection{Visualization of Temporal Attention}
To further illustrate the effect of temporal attention, we calculate all the temporal attention vectors on the testing dataset. As is shown in Figure~\ref{fig:cluster}, the last plot with the title \emph{Overall Distribution} implies that, in general, the news reported recently has more impact on the current stock trend than those reported earlier. 

Next, we adopt the spectral cluster method to generate five clusters representing different patterns of temporal attention, as is shown in the first five plots in Figure~\ref{fig:cluster}. We can find that despite the overall distribution, in different contexts with different news contents, the attention distribution can be quite diverse as well as dynamic. The informative news even in a few days ago can also express their influences in the temporal attention layer.
%Instead depended on the different sequential context, the news source in each day have the chance to be valued most.

\subsection{Effects of Self-paced learning}

To evaluate the effectiveness of the self-paced learning (SPL) in our framework, we compare the training process using the SPL with that using the standard training process. Both the convergence speed and the final result are considered as the comparison metric. As is shown in Figure~\ref{fig:curve}, at the beginning, the learning speed for SPL is slower than the one without using it, probably because SPL could neglect some difficult samples at early stages. However, as the round of epochs increases, the testing accuracy of the model using SPL gradually outperforms the one without using it, and eventually converge to a better result.

\subsection{Overall Performance Experiments}

\subsubsection{Classification Accuracy Result}

In the experimental setting of a tri-label classification problem, as the three label is approximately evenly split, we choose accuracy, which means the proportion of true results among the total number of testing samples, as the evaluation metric. The final results of all the methods are shown in Figure~\ref{fig:result}, in which each bar indicates the average accuracy of the testing dataset. Our proposed framework is able to gain the best accuracy result among all the baseline methods.

\subsubsection{Market Trading Simulation}

To further evaluate the effectiveness of our proposed framework, we conduct a back-testing by simulating the stock trading for approximately one year, from May 2016 to March 2017. 

Our estimation strategy conducts the trading in the daily frequency. At the beginning of each trading date, the model will give each stock a score based on the probability to have a rising trend minus the probability to have a declining trend. Based on these scores, a straightforward portfolio construction strategy called top-K selects $K$ stocks with the highest scores to construct a new portfolio for the next trading day. The selected $K$ stocks are evenly invested according to their \emph{OpenPrice}s of the next trading day. To approximate the real-world trading, we also consider a transaction cost of $0.3\%$ for each trading. We then calculate the average return on the stock market by evenly holding every stock as the baseline, indicating the overall market trend.

\begin{figure}[ht]
\centering   
\includegraphics [width=0.5\textwidth]{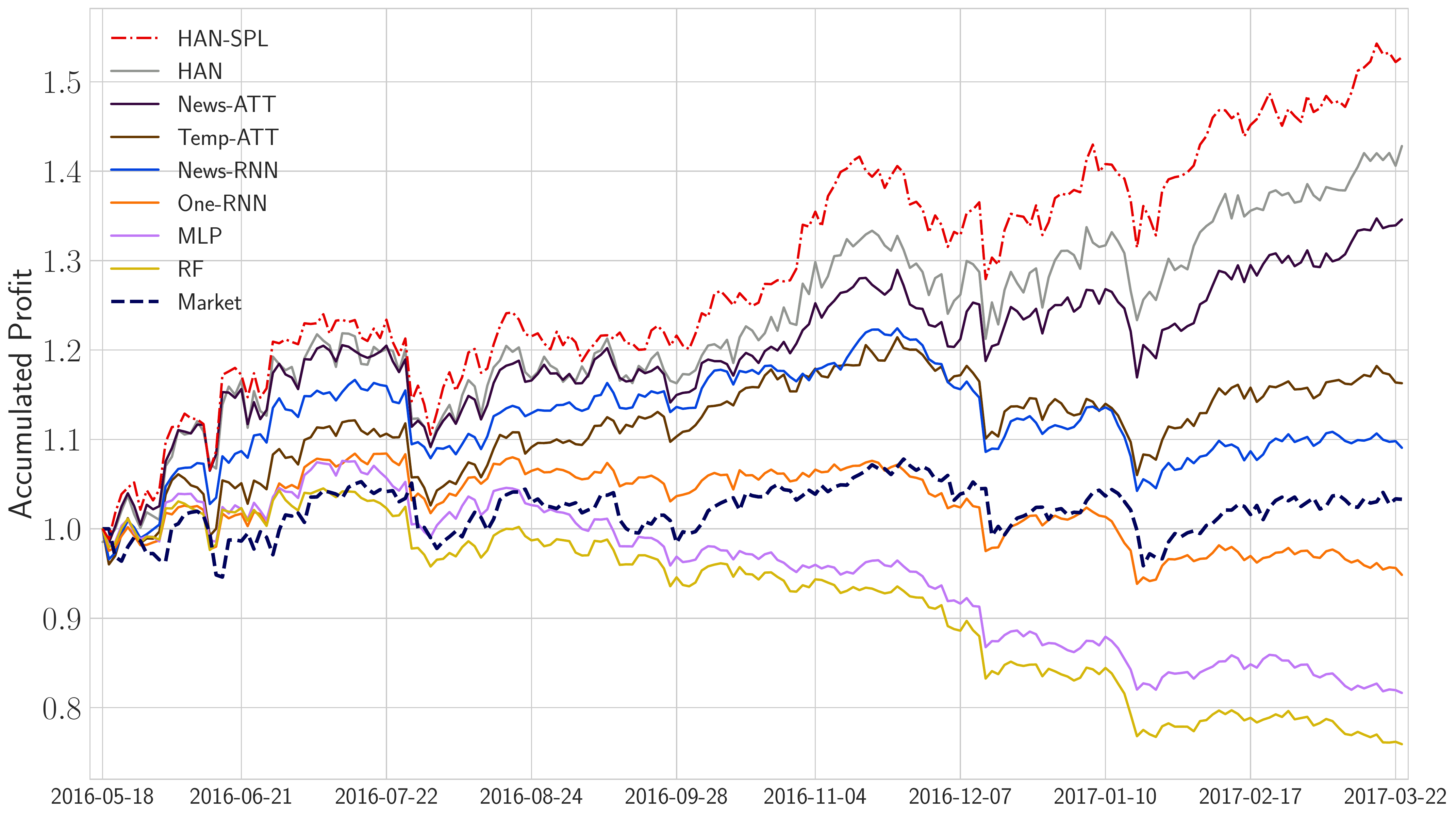} 
\caption{The cumulative profit curve of different methods with the portfolio of choosing top 40 stocks.} 
 \label{fig:cumulative} 
\end{figure}

To evaluate the performance of each prediction method, we use the annualized return as the metric, which is the cumulative profit per year. In reality, investors always choose multiple stocks to avoid risks. Therefore, we evenly invest on top $K$ stocks, where $k$ is set as $20$, $40$, $60$ and $80$, respectively, to compare the results in Figure~\ref{fig:annu}. In general, the highest profits should be obtained by selecting the most top stocks. However, when the $K$ is $20$, the improvement is not that obvious. %, while the annualized return of some methods is even lower. 
It is because that selecting few stocks leads to high turnover rate, which causes high transaction cost and offsets the profits. The detailed comparison and analysis of each algorithm will be elaborated in the next section. 

We also plot a cumulative profit curve for each method with $K$ as $40$. As is illustrated in Figure~\ref{fig:cumulative}, the cumulative profit results are consistent with the Overall Performance Analysis (section 5.4), and our proposed framework is able to gain the best profit result among all the baseline methods.

\subsection{Performance Discussion}
As is illustrated in Figure~\ref{fig:result} and Figure~\ref{fig:annu}, the tendency of the accuracy result and the annualized return result are consistent among different methods. Furthermore, the profit curves in Figure~\ref{fig:cumulative} demonstrated that the profit gaps between different methods almost do not change with time. Our proposed framework, HAN-SPL, can achieve the best accuracy of $0.478$ (Figure~\ref{fig:result}) and can obtain the highest profits in all different settings (Figure~\ref{fig:annu}). In particular, the highest annualized return $0.611$ are obtained while investing on top $40$ stocks, which is a remarkable improvement compared to the Market performance of $0.04$. We further analyze the results of different methods to discuss why our proposed framework can take effects. The discussions are as follows:

\noindent\textbf{Discussion on the RNN setting:}
The performances of MLP and RF are apparently worse than the others, probably because the input to these two methods are not organized in sequential contexts. This result to some extent indicates the significance to use RNN, which processes the news in the sequential context.

\noindent\textbf{Discussion on the bi-directional GRU setting:}
In our experiment setting, the News-RNN and One-RNN share the similar model structure, except that News-RNN uses a bi-directional GRU while One-RNN uses a one-directional GRU. From the figures, we can see that the News-RNN outperforms the One-RNN, showing the bi-directional setting, which can utilize the information from both past and future for prediction, effectively increases the performance.

\noindent\textbf{Discussion on the attention mechanisms:}
As we can see from the figures, the two baseline models with one attention layer, Temp-ATT and News-ATT, can achieve a better result than that of News-RNN, which has no attention layer, indicating the effectiveness to distinguish diverse influences of different news and dates. %And the better performance result of News-ATT compared to that of Temp-ATT implies that in our setting news-level attention is more efficient than the temporal-level attention. 

\noindent\textbf{Discussion on our proposed HAN framework:}
When we combine the two attention layers, our HAN can achieve even higher performance than all above-mentioned models, indicating the effectiveness of our overall model structure. In addition, the HAN model trained by self-paced learning (HAN-SPL) outperforms HAN with the standard training process, which shows the importance of effective and efficient learning. In Figure~\ref{fig:cumulative}, our HAN-SPL significantly outperforms all other methods for all the test time, which shows our proposed framework is indeed effective in the problem of stock trend prediction.

\section{Conclusion and Future Work}\label{sec:conclusion}
In this paper, we pointed out three principles for news-oriented stock trend prediction, including \textit{sequential context dependency}, \textit{diverse influence} and \textit{effective and efficient learning}, by imitating the learning process of human. Based on these principles, we proposed a new learning framework, particularly a Hybrid Attention Network (HAN) with self-paced learning mechanism, for stock trend prediction from online news. Extensive experiments on real stock market data demonstrated that our proposed framework can result in significant improvement in terms of the accuracy of stock trend prediction. Meanwhile, through back-testing, our framework can produce appreciable profits, with highly increased annualized excess return, in a one-year round trading simulation. 

In the future, beyond modeling the sequence of news related to one stock, we plan to further leverage the relationship between news related to different individual stocks, according to their industrial connections in real world. Moreover, we will investigate how to integrate the news-oriented approach with technical analysis for more accurate stock trend prediction.

% \section{Acknowledgements}
% This work was supported by the National Key Research and Development Program under the Grant No.  2016YFB1000105, the Natural Science Foundation of China (Grant No. 61725201 61421091, 61528201), and the Microsoft-PKU Joint Research Program. Xuanzhe Liu and Jiang Bian are corresponding authors of this work.

\bibliographystyle{abbrv}
\bibliography{wsdm}

\end{document}